\documentclass[prl,twocolumn,showpacs,floatfix,amsfonts]{revtex4}
\usepackage{graphicx,graphics,color,epsfig}% Include figure files
\usepackage{bm}
\usepackage{amsmath}
\usepackage{amssymb}

\newcommand{\beqa}{\begin{eqnarray}}
\newcommand{\eeqa}{\end{eqnarray}}

\begin{document}
\preprint{}

\title{Klein spin model ground states on general lattices}

\date{Received \today }

\author{Zohar Nussinov}
\affiliation{Dept. of Physics, Washington University, St. Louis,
MO 63160}

\begin{abstract}
We prove that in short range Klein spin models on general lattices, 
all ground states are of the dimer type- each fundamental 
plaquette must host at least one singlet. These ground states
are known to rigorously exhibit high dimensional fractionalization.
When combined with a recent theorem, 
this establishes that Klein spin models
exhibit topological order on the pyrochlore and checkerboard lattices.
\end{abstract}

\maketitle

\section{Introduction}

In this article, we illustrate
that in general short range Klein type spin models,
{\em all} ground states must be of the 
valence bond type. This will allow us to establish  
high dimensional fractionalization 
in these systems. We start with
a concise definition of the Klein 
model Hamiltonian \cite{Klein82}. We first divide the system into fundamental
units $\Box$. With the sum of all spins in $\Box$
denoted by $S_{\Box}$, and $S_{\max} = S \times n_{\Box}$
with $S$ the size of the spin at every site
and $n_{\Box}$ the number of the spins in $\Box$, 
the Klein model Hamiltonian is given by \cite{Klein82}
\begin{eqnarray}
H = J \sum_{\Box} P_{S_{\Box} = S_{\max}}~~~~~(J>0).
\label{Ham}
\end{eqnarray}
Here, $P_{S_{\Box}=S_{\max}}$ the projection 
operator onto the space with maximal
spin. For instance, for a spin $S=1/2$
system on a square lattice, $\Box$ is the elementary four site 
plaquette; here, we project onto the state of spin 
$S_{\max} = 4 \times \frac{1}{2} =2$.
Explicit forms for Klein models for
the hexagonal, \cite{Chayes89} square, \cite{rbt} decorated and general other
lattices \cite{raman}, 
extension of \cite{rbt} to $S=1$ \cite{NR}, 
and a slightly comprehensive study of pyrochlore lattices
\cite{NBNT} are available. In \cite{rbt} it was first shown how 
this may lead to fractionalization and effective dimensional reduction- 
a theme which was extended in \cite{NBNT} where
spinons could propagate in finite size volume fractions of the lattice.
The central thesis of many of these results is that
the pertinent ground states are of the dimer type. I.e., 
all ground states are superpositions 
of singlet product states,
\begin{eqnarray}
\label{singlet}
| \psi_{w} \rangle = \Big[ \bigotimes_{ij \in {\cal{W}} } |S_{ij} \rangle \Big]
\bigotimes |f \rangle.
\label{ws}
\end{eqnarray}

Here, $|S_{ij} \rangle$ is a singlet state formed 
between the two spins found at sites $i$ and $j$. In Eq.(\ref{ws}), 
${\cal{W}}$ denote a partition of sites in plaquettes on the lattice
into such singlet dimers such that, at least, one dimer
appears in every plaquette. Clearly, on some choices of plaquettes on 
various lattices,
such as the triangular lattice, if the elementary plaquette
may be chosen to be any spin and all of its nearest neighbors,
the constraint of one dimer per plaquette allows for many
spins to not be in any dimer singlet states. \cite{Chayes89} This extra
freedom that several remaining spins on such lattices
may enjoy is specified by $|f \rangle$. On the lattices
that we will focus on here (the square, checkerboard, and pyrochlore
lattices), no such freedom is found (the factor $|f \rangle$ does not appear).

While it is obvious that if any elementary plaquette hosts
at least one singlet dimer, $S_{\Box} < S_{\max}$ and thus
is a ground state, the converse is not as obvious. In \cite{Chayes89},
it was shown how the Klein spin model on the hexagonal lattice, 
two leg ladders, and on
octagonal diamond lattices needed to be of superpositions of
only dimer states of the form of Eq.(\ref{singlet}).   
We now prove that this result is universal and holds 
on a far greater variety of lattices. This proof 
is the central result of this article. 

\section{The single plaquette ground states}

Henceforth, we focus on $S=1/2$ spin systems. 
On any lattice, a single plaquette state is completely
symmetric under all permutations if and only if it lies in the sector of 
total spin $S_{\Box} = S_{max}$. As each permutation can be expressed
as a product of pair permutations, it follows
that 
\begin{eqnarray}
P_{S_{\Box} = S_{\max}} | \phi \rangle =0
\label{zeroc}
\end{eqnarray}
iff $|\phi \rangle$ can be written 
as a sum of plaquette state with
at least one (antisymmetric) singlet state
between two sites. Any wavefunction which 
is orthogonal to all plaquette wavefunctions
having, at least, one singlet state
must be completely symmetric in 
all of the spins and therefore lies entirely
in $S_{\Box}=S_{\max}$ sector. 
Far more demanding 
symmetry/antisymmetry restrictions
of the wavefunction on the entire lattice formed the corner
stone of the analysis of \cite{Chayes89}. 
We will now show that the above single plaquette result 
leads to the form of the ground state on the entire lattice.

\section{Ground states on the entire lattice}
We now generalize the single plaquette result to
the entire lattice.
As (i) the projection operator on any plaquette ($\Box_{i}$)
is non-negative definite, 
\begin{eqnarray}
P_{S_{\Box} = S_{\max}}^{2} = 
 P_{S_{\Box} = S_{\max}},
\label{square+}
\end{eqnarray} (ii) states with 
one dimer per plaquette have zero energy
saturate consequent the low energy bound ($E=0$)
and are thus ground states (yet not 
obviously the sole ground states),
and as (iii) the Hamiltonian is a sum over
non-negative definite projection operators
over each plaquette, we have
that in any ground state $| \psi \rangle$
\begin{eqnarray}
0 = \langle \psi| H | \psi \rangle \ge \langle \psi | P_{S_{\Box_{i}} 
= S_{\max}}| \psi \rangle.
\end{eqnarray}
Here, $\Box_{i}$ is any chosen plaquette $i$.  
As $\langle \psi | P_{S_{\Box_{i}} 
= S_{\max}}| \psi \rangle \ge 0$,
this, of course, mandates that
$\langle \psi | P_{S_{\Box_{i}} 
= S_{\max}}| \psi \rangle = 0$
for all plaquettes $i$. Now, by Eq.(\ref{square+}),
\begin{eqnarray}
\langle \psi | P_{S_{\Box_{i}} 
= S_{\max}}| \psi \rangle = \langle \phi_{i} | \phi_{i} \rangle, \nonumber
\\ {\mbox{with}}~~ |\phi_{i} \rangle \equiv  P_{S_{\Box_{i}} 
= S_{\max}}| \psi \rangle.
\end{eqnarray}
If $\langle \psi |  P_{S_{\Box_{i}} 
= S_{\max}}| \psi \rangle =0$ then 
$| \phi_{i} \rangle = 0$. 
Thus, in the ground state we need to satisfy 
$| \phi_{i} \rangle =0$ for all plaquettes $i$. 

We next write the state $| \psi \rangle$ in a general 
form in the complete orthonormal $\bigotimes_{r \in \Lambda} \sigma_{r}^{z}$
eigenbasis of the entire lattice ($\Lambda$) 
\begin{eqnarray}
| \psi \rangle = \sum_{\sigma^{z}_{j} \not \in \Box_{i}} 
\bigotimes_{j} | \sigma_{j}^{z} \rangle \sum_{\sigma^{z}_{k \in \Box_{i}}} 
\chi(\sigma_{j}^{z}; \sigma^{z}_{k \in \Box_{i}}) \bigotimes_{k \in \Box_{i}} 
| \sigma^{z}_{i} \rangle.
\end{eqnarray}
We now write the conditions which 
we found before
\begin{eqnarray}
0 =  P_{S_{\Box_{i}} 
= S_{\max}}| \psi \rangle \nonumber
\\ =  \sum_{\sigma^{z}_{j} \notin \Box_{i}} 
\bigotimes_{j} | \sigma_{j}^{z} \rangle \sum_{\sigma^{z}_{k \in \Box_{i}}} 
 P_{S_{\Box_{i}} = S_{\max}} \chi(\sigma_{j}^{z}; 
\sigma^{z}_{k \in \Box_{i}}) \bigotimes_{k \in \Box_{i}} 
| \sigma^{z}_{i} \rangle.
\end{eqnarray}

Next, let us define
\begin{eqnarray}
|\psi_{\Box_{i}} \rangle^{\sigma^{z}_{j} \notin \Box_{i}} \equiv 
\sum_{\sigma_{k}^{z} \in \Box_{i}} \chi(\sigma_{j \not \in \Box_{i}}^{z};
\sigma^{z}_{k \in \Box_{i}}) \bigotimes_{k \in \Box_{i}} |\sigma_{i}^{z} 
\rangle.
\end{eqnarray} 

If $P_{S_{\Box_{i}} = S_{\max}}| \psi_{\Box_{i}} 
\rangle^{\sigma^{z}_{j} \notin \Box_{i}} =0$ then,
according to our earlier proof of the last section
(following Eq.(\ref{zeroc}))
concerning the single plaquette wavefunctions,
$| \psi_{\Box_{i}} 
\rangle^{\sigma^{z}_{j} \notin \Box_{i}}$ must be a 
superposition of states each of which has, at least,
one singlet. As this holds for all
plaquettes $\Box_{i}$, we must have 
\begin{eqnarray}
|  \psi_{\Box_{i}} 
\rangle^{\sigma^{z}_{j} \notin \Box_{i}} =
\sum_{\alpha \beta \in \Box_{i}} 
c_{\alpha \beta}^{\sigma^{z}_{j} \notin \Box_{i}}
|\omega_{\alpha \beta}\rangle^{\sigma^{z}_{j} \notin \Box_{i}}   \nonumber
\\ ~{\mbox{where}}~ {\cal P}_{\Box_{i}}^{\alpha \beta} | \omega_{\alpha \beta} 
\rangle^{\sigma^{z}_{j} \notin \Box_{i}} = 
- | \omega_{\alpha \beta} \rangle^{\sigma^{z}_{j} \notin \Box_{i}},
~~ \alpha,
\beta \in \Box_{i}.
\end{eqnarray}
Here, ${\cal{P}}_{\Box_{i}}^{\alpha \beta}$ 
is the operator permuting the two sites $\alpha, \beta \in \Box_{i}$.
Every plaquette ($\Box_{i}$) must therefore have
at least one intra-plaquette singlet dimer (or
superpositions of states thereof). 
This concludes our proof. 

\begin{figure}[t!]
\vspace*{-0.5cm}
\includegraphics[angle=0,width=9cm]{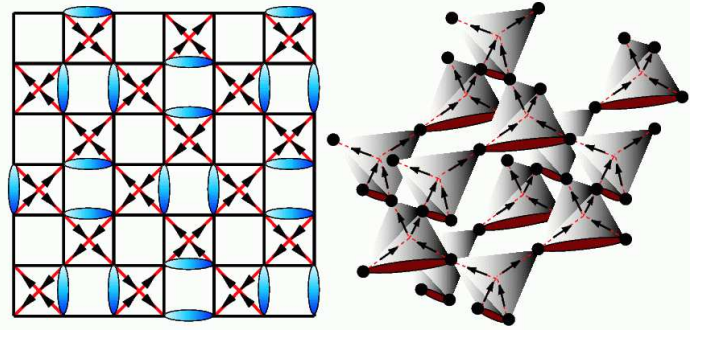}
\vspace{-1.5cm}
\caption{ (Color online.) 
From \cite{NBNT}. 
Highly regular ground states on the checkerboard and pyrochlore 
lattices. The ovals denote singlet dimer states. The arrows denote the 
representations of these dimer states within the six--vertex model (see 
Fig.(\ref{fig:pr1})) wherein on
each plaquette (tetrahedron or crossed square) 
a dimer connects the bases of the two arrows 
going to the center of the plaquette
on which the dimer is found.}
\label{PR}
\bigskip 
\end{figure}

On general lattices, all ground states are superspositions of states 
of the form of Eq.(\ref{ws}). All ground states 
are superpositions of states which have, at least,
one singlet on every plaquette. For $S=1/2$ 
Klein models on the square, pyrochlore, or checkerboard lattices
endowed with periodic boundary conditions,
the most general states satisfying this requirement
are of the form
\begin{eqnarray}
| \psi \rangle = \sum_{\alpha \beta \in \Box_{i}}  w_{\alpha \beta}
\bigotimes_{\alpha \beta \in \Box_{i}} \frac{1}{\sqrt{2}} 
\big[ | \uparrow_{\alpha} \downarrow_{\beta} \rangle -
| \downarrow_{\alpha} \uparrow_{\beta} \rangle \big].
\end{eqnarray}
Here, each lattice site appears in a singlet. 
An example of one such ground state is shown in Fig.(\ref{PR}).

Key steps in the above proof are (i)
the basis states $\bigotimes_{r} | \sigma^{z}_{r} \rangle$
are linearly independent; this allowed us to impose the condition
$ | \phi_{i} \rangle = P_{S_{\Box_{i}} = S_{\max}}| \psi_{\Box_{i}} 
\rangle^{\sigma^{z}_{j} \notin \Box_{i}} =0$ 
for {\em any} $\{ \sigma^{z}_{j \notin \Box_{i}} \}$,
(ii) The projection operator squared is equal
to itself, Eq.(\ref{square+}); 
this led to the condition $| \phi_{i} \rangle =0$
for all plaquettes $i$.

A decoration procedure readily allows 
for a demonstration of a
gap between the ground and excited states. \cite{raman}
Work in progress reaffirms this result for
all lattices.

\section{Topological order on the pyrochlore and checkerboard lattices}

\begin{figure}[t!]
\vspace*{-0.5cm}
\includegraphics[angle=90,width=6cm]{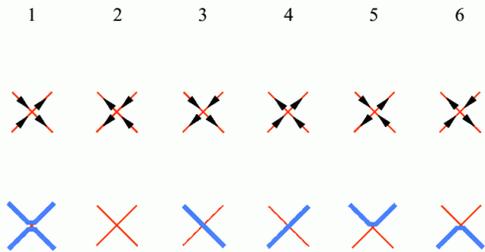}
\vspace{-1.5cm}
\caption{ (Color online.)
From \cite{NBNT} (Color online.) 
Standard representation of six--vertex states in terms of 
lines \cite{Baxter}. There are six possible dimer state
on each four site plaquette. In terms of the arrows referred to 
in Fig.(\ref{PR}), these correspond to six-vertex states
as shown here. Every line is composed of links whose arrows flow
to the right in the vertex representation.}
\label{rules}
\end{figure} 

\begin{figure}[t!]
\vspace*{-0.5cm}
\includegraphics[angle=90,width=9cm]{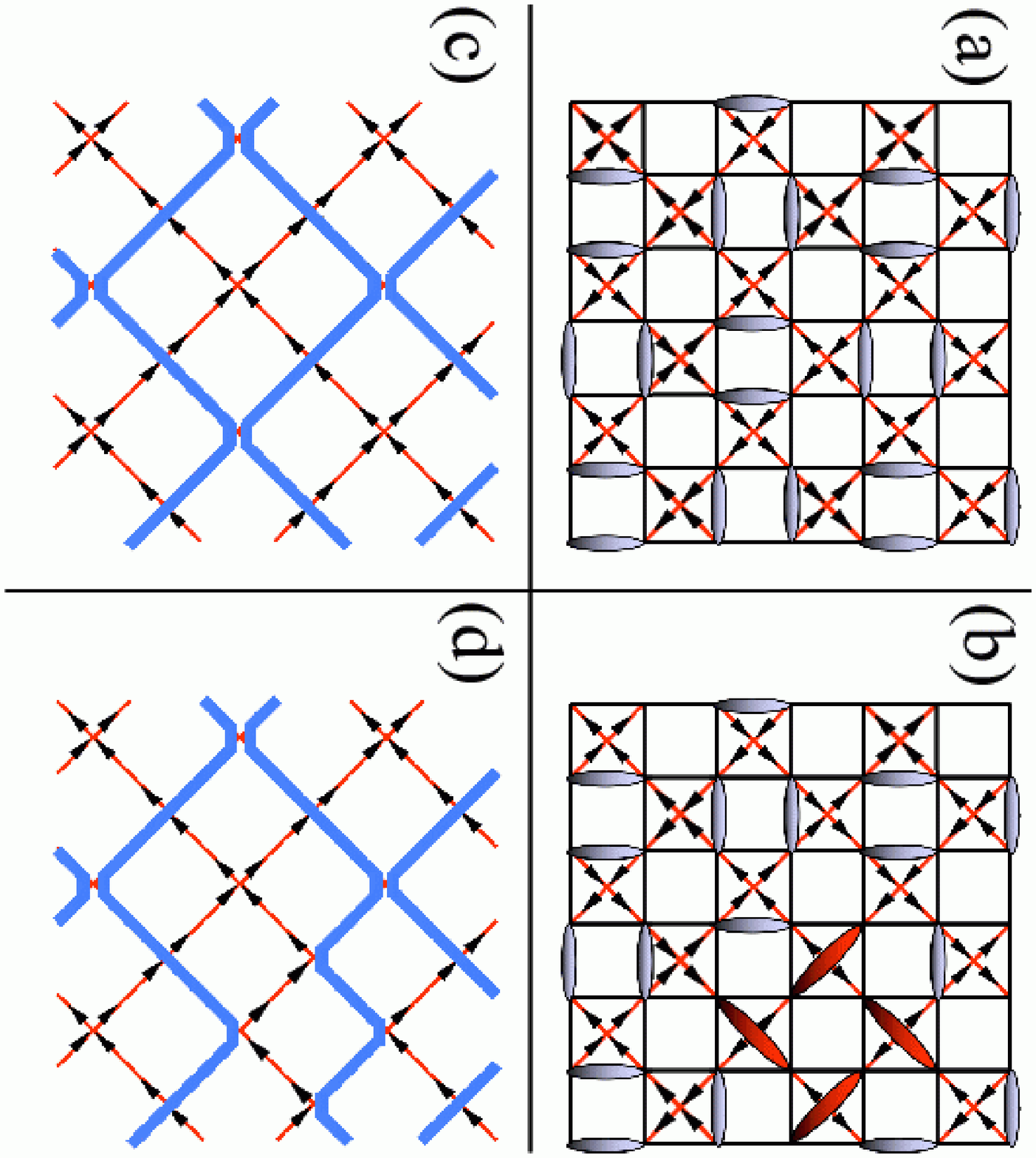}
\vspace{-1.5cm}
\caption{ (Color online.)  From \cite{NBNT}.
Left: representation of six--vertex states in a highly regular 
dimer configuration (a) in terms of lines (c) [see Fig.(\ref{rules}.)] 
Joining the vertex coverings 
according to the standard prescription  \cite{Baxter} allows 
to label all permitted ground states. Right: an elementary local 
spin interaction process, leading to a new dimer configuration (b), 
corresponds in the line representation (d) to flipping one of the line 
corners.}
\label{fig:pr1}
\end{figure}
To explicitly flesh out some of the more
abstract concepts discussed hitherto, we
examine the Klein model on the pyrochlore lattice.
The Klein Hamiltonian on the pyrochlore and
checkerboard lattices is a sum of nearest neighbor 
interactions augmented by ring exchange terms. Here, 
the Hamiltonian is of the form \cite{NBNT}
\begin{eqnarray}
H = J_{1} \sum_{\langle ij \rangle, \alpha} 
{\vec S}_i^{\alpha} \cdot 
{\vec S}_j^{\alpha} + J_{2} 
\sum_{\alpha} 
[({\vec S}_i^{\alpha} \cdot 
{\vec S}_j^{\alpha})
({\vec S}_k^{\alpha} \cdot 
{\vec S}_l^{\alpha})+\nonumber
\\ 
({\vec S}_i^{\alpha} \cdot 
{\vec S}_l^{\alpha})
({\vec S}_j^{\alpha} \cdot 
{\vec S}_k^{\alpha}) 
+ ({\vec S}_i^{\alpha} \cdot 
{\vec S}_k^{\alpha})
({\vec S}_j^{\alpha} \cdot 
{\vec S}_l^{\alpha})], \nonumber
\label{esh}
\end{eqnarray}
where $\langle ij \rangle$ denotes all pairs of sites in the tetrahedron
$\alpha \equiv ijkl$. The Klein model is realized when $J_2 = 4J_{1}/5$. On the 
checkerboard lattice, $\alpha$ denotes each  tetrahedral unit of
the pyrochlore lattice or crossed plaquette 
of the checkerboard lattice with cross--coupling interactions.
The first term is equivalent to both nearest-- and next--neighbor 
Heisenberg interactions of strength $J_{1}$. 
The Heisenberg interaction on the tetrahedral units (crossed
plaquettes of the checkerboard lattice).
A detailed study (wherein fractionalization and criticality 
were established) of the Klein model that results 
and fluctuations about it was undertaken
in \cite{NBNT}. Some highlights of this study are shown 
in Figs.(\ref{PR},\ref{rules},\ref{fig:pr1}).
As highlighted in Fig.(\ref{fig:pr1}), each dimer 
configuration corresponds to a system of continuous lines in the six-vertex 
representation. \cite{NBNT}, \cite{Baxter}, \cite{Lieb} 
  
An operational definition of zero temperature Topological 
Quantum Order (TQO) \cite{wenbook} in degenerate systems goes as follows:
Given a set of $N$ orthonormal ground states $\{|g_{\alpha} \rangle \}_{\alpha=1,..., N}$, TQO exists iff for any bounded operator $V$ with compact 
support (i.e. any quasi-local operator),
\begin{eqnarray}
\langle g_{\alpha} | V | g_{\beta} \rangle = v \delta_{\alpha \beta} + c,
\label{TQO}
\end{eqnarray}
where $v$ is a constant and $c$ is a correction that is either zero or
vanishes exponentially in the thermodynamic limit. \cite{kitaev, NO}
[Thus, in such systems, only
non-local ``topological'' quantities can become items of interest.] 
Relying on our demonstration that the Klein model only allows
for singlet covering ground states (and superpositions thereof)  
and that a finite spectral gap appears, topological
order on Klein spin systems on the pyrochlore and checkerboard lattices
follows. To see this, we note that, e.g., on
a checkerboard lattice in the sector
of a fixed number of lines, imposed by boundary conditions,
all states may be linked to each other by a sequence of 
local symmetry operations. Such a local symmetry operation 
is shown in Fig.(\ref{fig:pr1}). 
This emergent local ($d=0$ \cite{BN}) gauge-like symmetry
operation in a system with a gap mandates topological
quantum order as proved in \cite{NO}:
Glancing at Eq.(\ref{TQO}), we note that 
as no local symmetry can be
broken (Elitzur's theorem), \cite{BN} \cite{Elitzur} 
all diagonal elements of a non-gauge invariant observable $V$ vanish 
while all gauge invariant observable $V$ are the same regardless
of the state $\alpha$; the presence of the gap mandates that 
the off-diagonal terms of Eq.(\ref{TQO}) vanish. 
Even though, in principle, local observables such 
$V = (\vec{S}_{i} \cdot \vec{S}_{j}$) depend on whether or not 
a dimer exists between sites 
$i$ and $j$ in a given pure dimer state, e.g. 
$\langle \phi| \vec{S}_{i} \cdot \vec{S}_{j} 
| \phi \rangle = -1/4$ in a state $| \phi \rangle$ having a singlet
dimer between sites $i$ and $j$, such an operator unfortunately
cannot attain a non-zero value once infinitesimal 
symmetry restoring perturbations are introduced. This is similar to what
occurs in a $Z_{2}$ gauge theory given by  
the action $S = - K \sum_{\Box} U_{ij} U_{jk} U_{kl} 
U_{li}$ with the product of gauge fields $U_{ij} = \pm 1$ taken around the 
sides of a plaquette. (The gauge field $\{U_{ij} \}$ resides
on the link between sites $i$ and $j$.) Here,
$U_{ij}$ is different for different $Z_{2}$ gauge configurations.
However, by Elitzur's theorem $\langle U_{ij} \rangle =0$. \cite{Elitzur}
Similarly, in our case, $\langle V \rangle =0$ just as local 
observables vanish in gauge theory 
due to a generalized Elitzur's theorem applied to the non (local)
symmetry invariant
quantity $V$. \cite{BN}
An intuitive understanding may be gained by mapping Klein model ground states
onto $Z_{2}$ gauge theory ($K<0$) ground states; this (non-invertible) 
mapping proceeds by replacing a singlet dimer between 
sites $i$ and $j$ within the Klein model to a link with $U_{ij} =-1$
in a $Z_{2}$ gauge theory. 
A complete treatment entails the presence of a transverse like field
connecting the different dimer states (similar
to a transverse field in $Z_{2}$ gauge theories, e.g. \cite{hw}).
The result generally follows from symmetry 
considerations. \cite{NO}, \cite{BN} 
The line number of the 
six vertex representation is a conserved topological charge in these 
systems.

\section{Acknowledgments.} I am indebted to  
conversations with C. D. Batista and G. Ortiz.

\end{document}